\documentclass[preprint,aps]{revtex4}
\usepackage{graphicx}
\usepackage{dcolumn}
\usepackage{bm}

\begin{document}


\title{Stability beyond the neutron drip-line near the third peak of 
the r-process nucleosynthesis}

\author{M.M. SHARMA AND A.A. SALDANHA}
\address{Physics Department, Kuwait University, Kuwait 13060}


\begin{abstract}
We have investigated the nuclear shell effects at $N=126$ 
in the region of the third peak of the $r$-process nucleosynthesis 
within the framework of the relativistic mean-field theory using the 
Lagrangian model NL-SV1 with the vector self-coupling of $\omega$-meson. 
Our study encompasses even-even nuclei with $N=110-140$ in the isotopic 
chains of Hf ($Z=72$) down to Ba ($Z=56$). It is shown that the nuclear 
shell effects at $N=126$ remain strong even as one moves far away from 
the line of the $\beta$-stability. As the neutron drip line approaches
$N=126$, nuclei exhibit vanishingly small neutron separation energy. 
However, going beyond the neutron drip line, we observe an interesting 
feature in that some nuclei near $N \sim 132-134$ for the isotopic chains 
of $Z=62-68$ show enhanced neutron separation energy. This is especially
pronounced for the isotopes of Gd ($Z=64$) and Dy ($Z=66$). These nuclei 
exhibit the phenomenon of stability beyond the neutron drip line. 
Our analysis of the single-particle spectrum shows that this is 
engendered by the deformation assumed by these nuclei with the 
consequence that the neutron single-particle spectrum is pushed down 
in energy, thus leading to enhanced stability beyond the drip line.\\
\end{abstract}


\maketitle

\section{INTRODUCTION}
\label{Intro}

About half of nuclei heavier than Fe are synthesized in the process
of rapid neutron capture (r-process) (Burbidge {\it et al.} 1957, 
Hillebrandt 1978, Kratz {\it et al} 1993). 
Extremely neutron-rich nuclei with $\sim$ 10-30 neutrons 
away from the stability line are produced
in environments of high neutron densities and high temperatures, 
Due to extremely large isospin, these nuclei are highly unstable and 
are experimentally inaccessible, especially those in the heavy mass region. 
Undergoing a sequence of neutron capture accompanied by $\beta$-decays,
these extremely neutron-rich nuclei give birth to heavy elements in the 
nature. The r-process path passes through the magic numbers $N=$~50, 82 and 
126 at different mass values. The synthesis of nuclei around these 
magic numbers is reflected vividly in the known nuclear abundance
peaks around  $A \sim$ 80, 130 and 190, respectively. 

The shell effects at the magic numbers play a crucial role
in determining the r-process nuclear abundances  (Kratz {\it et al} 1993).
The question whether the shell effects near the r-process 
path are strong or do quench has become crucial to understanding 
the nucleosynthesis of heavy nuclei (Pfeiffer {\it et al} 2001).
It was argued for long that a quenching of the shell effects 
at $N=82$ near the r-process path was required in order to reproduce
$r$-process abundances in the second peak (Pfeiffer 1997).
This argument came primarily from the ability of the macroscopic-microscopic 
mass formula Extended Thomas-Fermi with Strutinsky Integral  (Strutinsky 1968)
built with quenching based upon Bogoliubov pairing (ETF-SI(Q)) 
(Pearson {\it et al.} 1996) to reproduce r-process abundances. 
This quenching has been termed as artificial, for it is based upon the 
extreme quenching due to the Skyrme force SkP (Dobaczewski {\it et al.}
1984). In contrast, microscopic theories such as the RMF theory with 
the Hartree-Bogoliubov approach have not shown any significant
weakening of the shell effects near the r-process path 
(Sharma and Farhan 2002). 

Recently, indications have emerged from analysis of several experimental
data that shell quenching which was invoked earlier is not required. Anomalous
behaviour of $2^+$ excitation energies in neutron-rich Cd isotopes was
cited as an evidence of shell quenching of $N=82$ (Dillman {\it et al.} 2002)
However, the recent work by Rodriguez {\it et al.} (2008) 
has shown  that there is no need to assume quenching of $N=82$ shell closure 
to explain the experimental finding. This is further corroborated by the 
recent experimental work (Jungclaus {\it et al.} 2007, Gorska {\it et al.}
2009), where results on level scheme of $^{130}$Cd show 
no evidence of quenching at $Z=48$. These works have confirmed the prediction
of the absence of quenching at $N=82$ shell closure for $Z=48$ and
neighbouring elements by one of the authors and a collaborator 
(Sharma and Farhan 2002) with the force NL-SV1 within the RMF theory. 
On the other hand, the shell effects at $N=82$ were shown to 
weaken only near the neutron drip line. In that work  (Sharma and Farhan
2002),  it was shown that force NL-SV1 based upon the vector self-coupling of 
$\omega$-meson (Sharma {\it et al.} 2000), which reproduces the shell effects 
in nuclei at the stability line is able to reproduce the available data on the 
waiting-point nucleus $^{80}$Zn  (Sharma and Farhan 2002). 

The shell effects at N=126 near the r-process path were also investigated 
(Farhan and Sharma 2006)
within the framework of the Relativistic Hartree-Bogoliubov
theory with the Lagrangian model NL-SV1 using spherical configuration of
nuclei. In contrast to the N=82 shell effects, it was shown that 
the shell strength at N=126 remains strong even in approaching the
neutron drip line. The spherical configuration of nuclei is appropriate
for nuclei in the vicinity of a shell closure especially when it remains
robust. However, away from the shell closure nuclei assume deformation.

In the present work, we have investigated the ground-state properties
of nuclei across the shell closure $N=126$ within the framework of
the deformed RMF theory. In Section 2, we present a brief description of
the formalism of the RMF theory. Section 3 provides the detailed results
of the study of the rare-earth nuclei spanning the shell closure
$N=126$. The final section provides the summary and conclusions.

\section{THE RELATIVISTIC MEAN-FIELD THEORY: FORMALISM}

The RMF approach (Serot and Walecka 1986) is based upon the Lagrangian density 
which consists of fields due to the various mesons interacting with 
the nucleons. The mesons include the isoscalar scalar $\sigma$-meson, 
the isovector vector $\omega$-meson and the isovector vector $\rho$-meson.
The Lagrangian density is given by:

\begin{equation}
\begin{array}{rl}
{\cal L} &=
\bar \psi (i\rlap{/}\partial -M) \psi +
\,{1\over2}\partial_\mu\sigma\partial^\mu\sigma-U(\sigma)
-{1\over4}\Omega_{\mu\nu}\Omega^{\mu\nu}+\\
\                                        
\ & {1\over2}m_\omega^2\omega_\mu\omega^\mu
+{1\over4}g_4(\omega_\mu\omega^\mu)^2
-{1\over4}{\bf R}_{\mu\nu}{\bf R}^{\mu\nu} +
 {1\over2}m_{\rho}^{2}
 \mbox{\boldmath $\rho$}_{\mu}\mbox{\boldmath $\rho$}^{\mu}
-{1\over4}F_{\mu\nu}F^{\mu\nu} \\
\                              
\ &  g_{\sigma}\bar\psi \sigma \psi~
     -~g_{\omega}\bar\psi \rlap{/}\omega \psi~
     -~g_{\rho}  \bar\psi
      \rlap{/}\mbox{\boldmath $\rho$}

     \mbox{\boldmath $\tau$} \psi
     -~e \bar\psi \rlap{/}{\bf A} \psi
\end{array}
\end{equation}
The bold-faced letters indicate the vector quantities. 
Here M, m$_{\sigma}$, m$_{\omega}$ and m$_{\rho}$ denote the nucleon-, 
the $\sigma$-, the $\omega$- and the $\rho$-meson masses respectively, 
while g$_{\sigma}$, g$_{\omega}$, g$_{\rho}$ and e$^2$/4$\pi$ = 1/137 are 
the corresponding coupling constants for the mesons and the photon,
respectively. 

The $\sigma$ meson is assumed to move in a scalar potential of the form:
\begin{equation}
U(\sigma)~={1\over2}m_{\sigma}^{2} \sigma^{2}~+~
{1\over3}g_{2}\sigma^{3}~+~{1\over4}g_{3}\sigma^{4}.
\end{equation}

This was introduced by Boguta and Bodmer (1977) in order to
make a substantial improvement in the surface properties of finite
nuclei. This Ansatz for the $\sigma$-potential has since become a 
standard and necessary ingredient for description of the properties of 
finite nuclei. 

In this work, we have employed the non-linear vector self-coupling of
the $\omega$-meson, in addition to the non-linear scalar potential
of Eq. (2). The coupling constant for the non-linear $\omega$-term
is denoted by $g_4$ in the Lagrangian (1). 

The field tensors of the vector mesons and of the electromagnetic
field take the following form:
\begin{equation}
\begin {array}{rl}
\Omega^{\mu\nu} =& \partial^{\mu}\omega^{\nu}-\partial^{\nu}\omega^{\mu}\\
\          
{\bf R}^{\mu\nu} =& \partial^{\mu}
                  \mbox{\boldmath $\rho$}^{\nu}
                  -\partial^{\nu}
                  \mbox{\boldmath $\rho$}^{\mu}\\
\                
F^{\mu\nu} =& \partial^{\mu}{\bf A}^{\nu}-\partial^{\nu}{\bf A}^{\mu}
\end{array}
\end{equation}

The mean-field approximation constitutes the lowest order of the
quantum field theory. Herein, the nucleons are assumed to move
independently in the meson fields. The latter are replaced by their
classical expectation values. The ground-state of the nucleus is 
described by a Slater determinant $\vert\Phi >$ of single-particle
spinors $\psi_i$ (i = 1,2,....A). The stationary state solutions $\psi_i$
are obtained from the coupled system of Dirac and Klein-Gordon equations.
The variational principle leads to the Dirac equation:
\begin{equation}
\{ -i{\bf {\alpha}} \nabla + V({\bf r}) + \beta [ m* ] \}
~\psi_{i} = ~\epsilon_{i} \psi_{i}
\end{equation}
where $V({\bf r})$ represents the $vector$ potential:
\begin{equation}
V({\bf r}) = g_{\omega} \omega_{0}({\bf r}) + g_{\rho}\tau_{3} {\bf {\rho}}
_{0}({\bf r}) + e{1-\tau_{3} \over 2} {A}_{0}({\bf r})
\end{equation}
and $S({\bf r})$ is the $scalar$ potential
\begin{equation}
S({\bf r}) = g_{\sigma} \sigma({\bf r})
\end{equation}
which defines the effective mass as:
\begin{equation}
m^{\ast}({\bf r}) = m + S({\bf r})
\end{equation}
The Klein-Gordon equations for the meson fields are time-independent
inhomogeneous equations with the nucleon densities as sources.
\begin{equation}
\begin{array}{ll}
\{ -\Delta + m_{\sigma}^{2} \}\sigma({\bf r})
 =& -g_{\sigma}\rho_{s}({\bf r})
-g_{2}\sigma^{2}({\bf r})-g_{3}\sigma^{3}({\bf r})\\
\         
\  \{ -\Delta + m_{\omega}^{2} \} \omega_{0}({\bf r})
=& g_{\omega}\rho_{v}({\bf r}) + g_4 \omega^3({\bf r}) \\
\                            
\  \{ -\Delta + m_{\rho}^{2} \}\rho_{0}({\bf r})
=& g_{\rho} \rho_{3}({\bf r})\\
\                           
\  -\Delta A_{0}({\bf r}) = e\rho_{c}({\bf r})
\end{array}
\end{equation}

For the case of an even-even nucleus with time-reversal symmetry, the
spatial components of the vector fields, 
\mbox{\boldmath $\omega$}, \mbox{\boldmath $\rho_3$} and
\mbox{\boldmath A} vanish. For the mean-field, the nucleon spinors
provide the corresponding source terms:
\begin{equation}
\begin{array}{ll}
\rho_{s} =& \sum\limits_{i=1}^{A} \bar\psi_{i}~\psi_{i}\\
\             
\rho_{v} =& \sum\limits_{i=1}^{A} \psi^{+}_{i}~\psi_{i}\\
\             
\rho_{3} =& \sum\limits_{p=1}^{Z}\psi^{+}_{p}~\psi_{p}~-~
\sum\limits_{n=1}^{N} \psi^{+}_{n}~\psi_{n}\\
\                    
\ \rho_{c} =& \sum\limits_{p=1}^{Z} \psi^{+}_{p}~\psi_{p}
\end{array}
\end{equation}
where the sums are taken over the valence nucleons only. Consequently, 
the ground-state of a nucleus is obtained by solving the coupled system 
of the Dirac and Klein-Gordon equations self-consistently.

\subsection{Axially deformed RMF}

We solve the Dirac equation  as well as the
Klein-Gordon equations  by expansion of 
the wavefunctions into a complete set of eigen 
solutions of an harmonic oscillator potential (Gambhir et al. 1990) 
In the axially symmetric case the spinors $f^\pm_i$ and $g^\pm_i$ are expanded 
in terms of the eigenfunctions of a deformed axially symmetric oscillator
potential 
\begin{equation}\label{eqn:2.47}
V_{osc}(z,r_\bot)=\frac{1}{2}M\omega^2_zz^2+\frac{1}{2}M\omega_\bot^2r_\bot^2.
\end{equation}
Taking volume conservation into account, the two oscillator frequencies
$\hbar\omega_\bot$ and $\hbar\omega_z$ can be expressed in terms of a
deformation parameter $\beta_0$.
\begin{equation}
\hbar\omega_z=\hbar\omega_{0}exp\Bigg(-\sqrt{\frac{5}{(4\pi)}}\beta_0\Bigg)
\end{equation}
\begin{equation}
\hbar\omega_\bot=\hbar\omega_{0}exp\Bigg(+\frac{1}{2}
\sqrt{\frac{5}{(4\pi)}}\beta_0\Bigg).
\end{equation}
The corresponding oscillator length parameters are given by
\begin{equation}
b_z=\sqrt{\frac{\hbar}{M\omega_z}}\qquad {\rm and} \qquad
b_\bot=\sqrt{\frac{\hbar}{M\omega_\bot}}.
\end{equation}
The volume conservation gives $b_\bot^2b_z=b_0^3$. The basis is now determined 
by the two constants $\hbar\omega_0$ and $\beta_0$, which 
are chosen optimally. 

The deformation parameter of the oscillator basis $\beta_0$ is 
chosen to be identical for the Dirac spinors and the meson fields.  
The deformation parameter $\beta_2$ is obtained from the calculated
quadrupole moments for protons and neutrons through
\begin{eqnarray}
Q=Q_n+Q_p&=&\sqrt{\frac{16\pi}{5}}\frac{3}{4\pi}AR_0^2\beta_2
\end{eqnarray}
with $R_0=1.2A^{1/3}$ (fm). The quadrupole 
moments is calculated as
\begin{eqnarray}
Q_{n,p}=\langle 2r^2P_2(cos\theta)\rangle_{n,p}
=2\langle 2z^2-x^2-y^2 \rangle_{n,p}.
\end{eqnarray}

\section{Details of calculations}

In this work we have employed the Lagrangian model NL-SV1 (Sharma et al. 2000)
with the inclusion of the vector self-coupling of $\omega$-meson in the
RMF Lagrangian. It was shown (Sharma et al. 2000) that inclusion of 
the vector self-coupling of $\omega$-meson provides an improved 
description of the shell effects in nuclei along the stability line.
The force NL-SV1  was developed with a view to improve the predictions of 
the ground-state properties of nuclei, such as binding energies, 
charge radii and isotopes shifts of nuclei along the stability line 
and far away from it. The parameters of the set NL-SV1  are given 
in Table I.

\begin{table}
\begin{center}
\caption {\label{table:}The Lagrangian parameters of the force NL-SV1 
(Sharma et al. 2000) used in the RMF calculations.}
\vglue0.7cm
\begin{tabular}{|l|r|}
\hline\hline
Parameters & NL-SV1~~~~\\
\hline\hline
~~~~$M$ ~~~~~&~~~~~939.0~~~~\\
~~~~$m_\sigma$~~~~~&~~~~~ 510.0349~~~~\\
~~~~$m_\omega$~~~~~&~~~~~ 783.0~~~~\\
~~~~$m_\rho$~~~~~&~~~~~   763.0~~~~\\
~~~~$g_\sigma$~~~~~&~~~~~  10.1248~~~~\\
~~~~$g_\omega$~~~~~&~~~~~ 12.7266~~~~\\
~~~~$g_\rho$~~~~~&~~~~~ 4.4920~~~~\\
~~~~$g_2$~~~~~&~~~~~  $-$9.2406~~~~\\
~~~~$g_3$~~~~~&~~~~~ $-$15.388~~~~\\
~~~~$g_4$~~~~~&~~~~~ 41.0102~~~~\\
\hline\hline
\end{tabular}
\end{center}
\end{table}

The input parameters required to carry out explicit numerical calculations
are: neutron pairing gap $\triangle_n$, the proton pairing gap $\triangle_p$
and the number of oscillator shells N$_F$ and N$_B$ of the fermionic
wavefunctions and meson fields, respectively. Both the fermionic and bosonic 
wavefunctions have been expanded in a basis of 20 harmonic oscillator shells 
in this work.

For pairing gaps, we have used the formula due to 
M\"oller and Nix (1992) as given by:
\begin{equation}
\begin{array}{cc}
\triangle_ {n} = 4.8 ~N^{-1/3}\\
\
\triangle_{p} = 4.8  ~Z^{-1/3}.
\end{array}
\end{equation}
An axially symmetric deformed configuration with reflection symmetry has been
assumed for nuclei. For each nucleus, RMF minimization has been sought
both in the prolate as well as in the oblate region of the 
deformation space. 

\section{RESULTS AND DISCUSSION}

As mentioned in the introduction, the nuclei corresponding to the
third peak in the r-process nucleosynthesis lie in the region of
the rare-earths in the periodic table. Especially, it concerns
the rare-earth nuclei beyond the magic number $N=126$ with elements
from $Z=56$ to $Z=72$. We have then investigated nuclei in this
region using the RMF theory with the Lagrangian model NL-SV1. 

\subsection{The binding energies}

The binding energy per nucleon of nuclides in various isotopic chains studied 
in this work is shown for the lowest energy minimum 
(ground state) in Fig.~1. Fig.~1(a) shows the binding energy per nucleon
for the chains of Hf ($Z=72$), Yb ($Z=70$), Er ($Z=68$) and Dy ($Z=66$,) 
and Fig.~1(b) shows those for the chains Gd ($Z=64$), Sm ($Z=62$), Nd ($Z=60$) 
and Ce ($Z=58$). The curves show a natural decline in the binding energy as 
the neutron number increases, for one is already treading farther and 
farther away from the line of $\beta$-stability as the neutron number 
increases. The difference in the relative binding energies for the 
various chains is also evident as the curves show a decrease as 
the isospin increases from one chain to the other (with a decrease 
in $Z$) for a given neutron number.

\begin{figure}[h*]
\vspace {0.5 cm}
\hspace{0.5cm}
\resizebox{0.85\textwidth}{!}{%
   \includegraphics{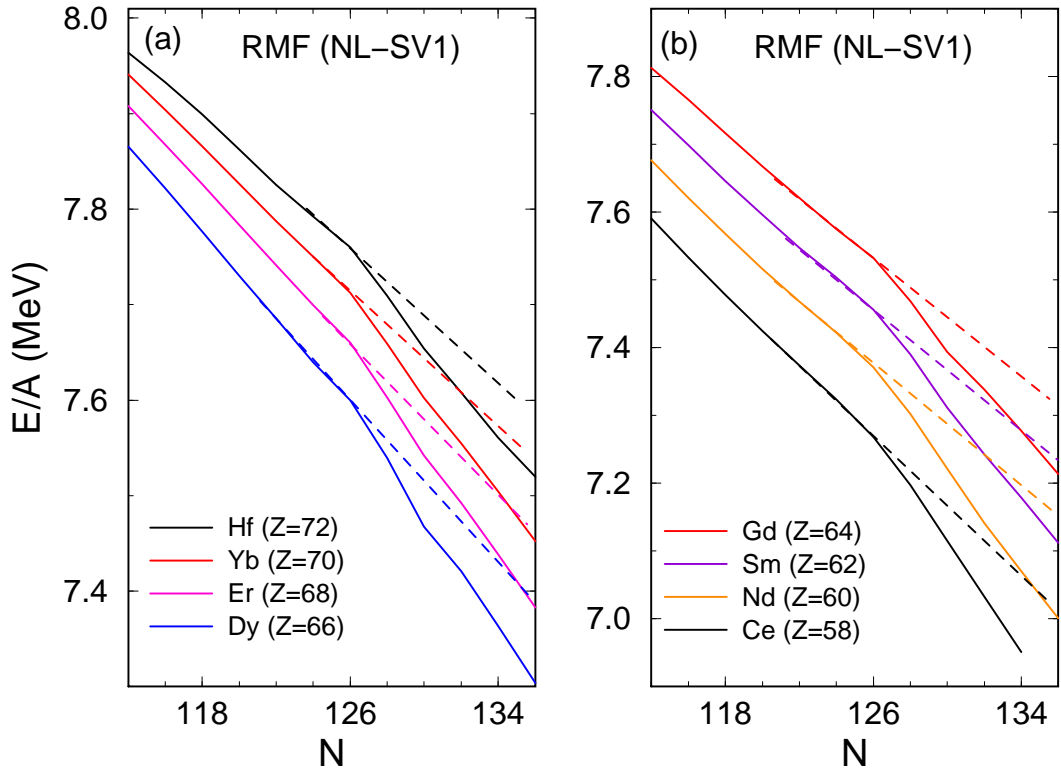}}
\vspace{0cm}       
\caption{The binding energies of nuclei obtained with the RMF theory
using the Lagrangian set NL-SV1 with the vector self-coupling of
$\omega$-meson. The kink at $N=126$ is seen clearly. The view is augmented 
by the dashed line at each curve, which has been drawn just to 
guide the eye.}
\label{fig:1}       
\end{figure}

The behaviour of the binding energies at the magic number $N=126$ is 
conspicuous.
The effect of the shell-closure is evident in the kink at $N=126$.
A downward and a larger slope above $N=126$ shows a rapid decrease in the 
binding energy for nuclides as compared to those below $N=126$. This change
in the slope is demonstrated by the dashed line which has been drawn 
to guide the eye. The divergence from the straight line (dashed) 
increases slightly especially for the isotopic chains 
below $Z=66$. This is due to the fact (as we will see in the 
later sections below) that nuclides with $N>126$ approach the 
drip line as one moves to the chains below $Z=66$. 
This indicates that for nuclides which are closer to the drip line, 
there is a rapid decline in the binding energy contribution due to 
the added neutrons.

\subsection{The deformation properties}

The nuclides in the isotopic chains studied in this work fall in the
region of the rare-earths as mentioned earlier.
It is well known that nuclei in this region assume rather 
strong deformations. Hence this work is intended to investigate
the properties of nuclei in this region in the deformed RMF theory.
As a result of axially deformed RMF 
minimizations, we have obtained deformation properties of nuclei. 
We show the ensuing quadrupole deformation $\beta_2$  of nuclides in 
various isotopic chains in Figs.~2 and 3. The figures are arranged in the 
descending order of atomic number $Z$ with a view to visualizing the 
effect of moving from the region closer to the stability line to 
that approaching the neutron drip line.

\begin{figure}[h*]
\vspace {0.5 cm}
\hspace{0.5cm}
\resizebox{0.70\textwidth}{!}{%
   \includegraphics{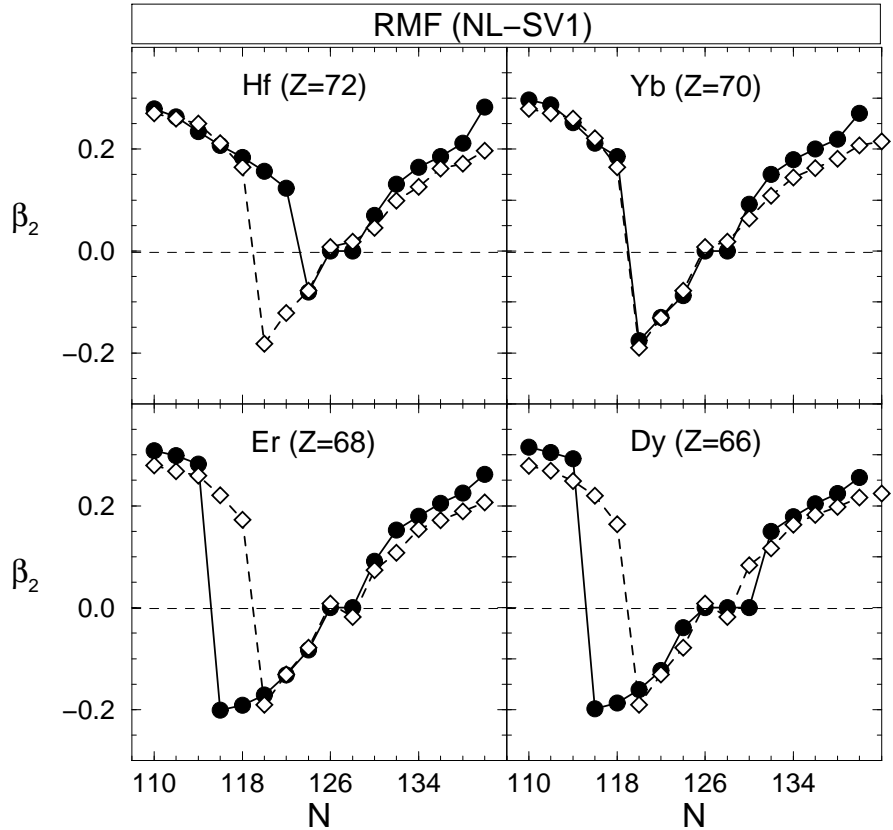}}
\vspace{1.0cm}       
\caption{The quadrupole deformation $\beta_2$ for the lowest 
minimum (the ground state) obtained in the RMF theory with 
NL-SV1 (full circles) for the isotopic chains $Z=66-72$. 
A comparison is made with the FRDM values (open diamonds).} 
\label{fig:2}       
\end{figure}

Fig.~2 shows the $\beta_2$ values for the isotopic chains of Hf ($Z=72$),
Yb ($Z=70$), Er ($Z=68$) and Dy ($Z=66$). Nuclides in these chains especially 
below the magic number $N=126$ are still away from the neutron drip line.
The neutron drip line in these chains is expected to 
lie far beyond the neutron number $N=140$.

Nuclides in all the isotopic chains in Fig. 2 exhibit an identical 
pattern of prolate
shape followed by a transition to an oblate shape, which is followed by
spherical shape near and at $N=126$, which is  followed by a prolate 
shape again for nuclides above $N=126$. 
Nuclides in these chains show a significant prolate 
deformation below $N=120$. For the Hf chain, several nuclides exhibit a 
prolate deformation which continues from $N=110$ until $N=122$. This is 
striking that Hf nuclides show more extended region of the prolate deformation 
in contrast to its neighbours with $Z=70$, $Z=68$ and $Z=66$. 
For the Hf chain, even the nuclide with $N=122$ which is so close to the 
magic number $N=126$ exhibits a prolate deformation. This is followed 
by a shape transition to a weakly deformed oblate shape at $N=124$ 
followed by sphericity for the magic nucleus $N=126$.

For the isotopic chains $Z=70$ and especially for $Z=68$ and $Z=66$, 
a fewer nuclei exhibit prolate shape as compared with $Z=72$. For $Z=70$, 
the prolate-to-oblate transition takes place at $N=120$, whereas
this is shifted further down to $N=116$ for $Z=68$ and $Z=66$. 
Consequently, the region of oblate shape increases to 3 nuclei for 
$Z=70$ and to 5 nuclei for $Z=68$ and $Z=66$ before achieving 
sphericity at N=126. However, for the chains with $Z=68-72$ in Fig. 2(a) 
only two nuclides, i.e., with $N=126$ and $N=128$ exhibit sphericity.
For the Dy ($Z=66$) chain, it increases to three nuclides which
are spherical near $N=126$. This would imply that the shell effect
is in no way becoming milder in going down from Hf ($Z=72$) to 
Dy ($Z=66$), i.e. the shell gap at $N=126$ retains its strong character
or rather it might even imply a somewhat strengthening of the 
shell effect which induces more sphericity in the vicinity of $N=126$.

For all the four chains in Fig.~2, nuclei above $N=128$ exhibit a prolate 
shape with increasing magnitude of deformation as one goes further away 
from the magic number $N=126$. This is natural, for the effect of the closed
shell gets diminished as one moves farther away from the shell closure.

A comparison of the $\beta_2$ values from the macroscopic-microscopic
mass formula FRDM (M\"oller et al. 1994) is made in Fig.~2. The FRDM
shows a remarkably similar feature such as transition from prolate to
oblate shape as the neutron number increases followed by a spherical 
shape near and at $N=126$, which is then followed by a prolate shape
as one moves beyond $N=126$. Incidentally, the number of nuclei exhibiting
spherical shape near $N=126$ with FRDM is nearly the same as with
NL-SV1. This may be due to the fact that the shell effects at
$N=126$ with FRDM are similar to those with NL-SV1.
This was already shown to be the case in the spherical relativistic 
Hartree-Bogoliubov approach (Farhan and Sharma 2006) with NL-SV1,
where a comparison was made with FRDM. 

The difference between the predictions of NL-SV1 and those of FRDM
are primarily about the point of transition from the prolate to the
oblate shape below $N=126$. For all the isotopic chains in Fig.~2, 
the prolate-oblate transition takes place at $N=120$. Thus, there
is a near constancy about it. In comparison, the transition point
is shifting to lower values of $N$ as one moves from $Z=72$ to
$Z=66$ with NL-SV1. This can be attributed to changing structural
factors as one traverses from one value of $Z$ to another. 

\begin{figure}[h*]
\vspace {0.5 cm}
\hspace{0.5cm}
\resizebox{0.70\textwidth}{!}{%
   \includegraphics{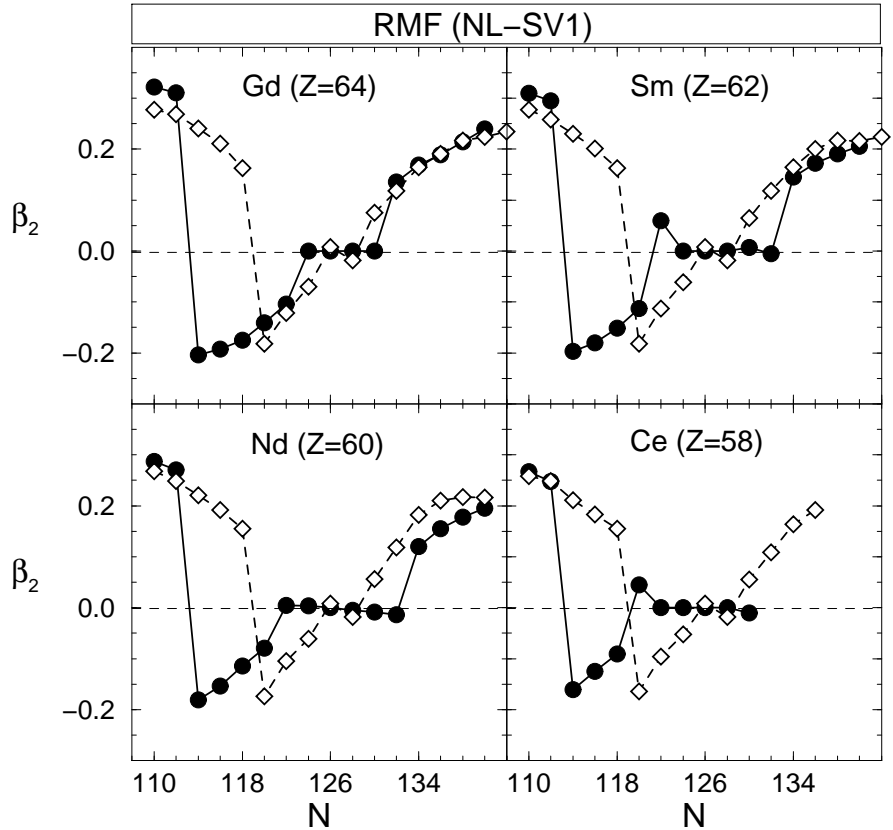}}
\vspace{1cm}       
\caption{The quadrupole deformation $\beta_2$ for the lowest 
minimum (the ground state) obtained in the RMF theory with 
NL-SV1 (full circles) for the isotopic chains $Z=58-64$. 
A comparison is made with the FRDM values (open diamonds).}
\label{fig:3}       
\end{figure}
 
The quadrupole deformation $\beta_2$ for the chains Gd ($Z=64$), Sm ($Z=62$),
Nd ($Z=60$) and Ce ($Z=58$) is shown in Fig.~3. The prolate-to-oblate 
transition is exhibited by all these chains as in Fig.~2. The transition
point, however, is shifted a few neutron number, i.e., it takes place at 
$N=114$ for all the chains with $Z=58-64$. Consequently, only a fewer
nuclides exhibit a prolate shape below $N=126$. As a result, more nuclei
exhibit oblate shape below $N=126$. At the same time, the number of nuclei
exhibiting a spherical shape near the magic number increases from 
4 ($N=124-130$) for $Z=64$ and to 6 ($N=122-132$) for $Z=62$. 
Incidentally, for these isotopic chains, magic nuclei and those beyond 
it are approaching the drip line closely as we will see below in the 
subsection on 2-neutron separation energies. However, in spite of the 
proximity of nuclei to the magic number $N=126$ for the chains Z=62-66 
non-magic nuclei near N=126 are not susceptible to a deformation. Also,
the number of spherical nuclei near $N=126$ is higher than those in
Fig.~2. This is due to the reason that in going to the lower Z values,
the strong shell character of the magic number $N=126$ is maintained
even as one approaches the neutron drip line.

A comparison of the $\beta_2$ values due to NL-SV1 with those
from FRDM shows that the basic feature of the shape transitions
in Fig.~3 is similar to that in Fig.~2. The FRDM values show
the prolate-oblate transition point at $N=120$ as in Fig.~2,
whereas it has shifted to $N=114$ for NL-SV1 for
all the chains in Fig.~3.

\subsection{The shape-coexistence near $N=126$}

The phenomena of shape-coexistence is well known in nuclides in
several parts of the periodic table. It arises due to interplay of deformation
and the consequent re-adjustment of single-particle levels. A large number
of nuclides in the isotopic chains we have investigated also exhibit the
phenomenon of the shape-coexistence of prolate-oblate shapes.
We show the nuclides exhibiting the shape-coexistence in this region 
in Figs.~4 and 5.

\begin{figure}[h*]
\vspace {0.5 cm}
\hspace{0.5cm}
\resizebox{0.80\textwidth}{!}{%
   \includegraphics{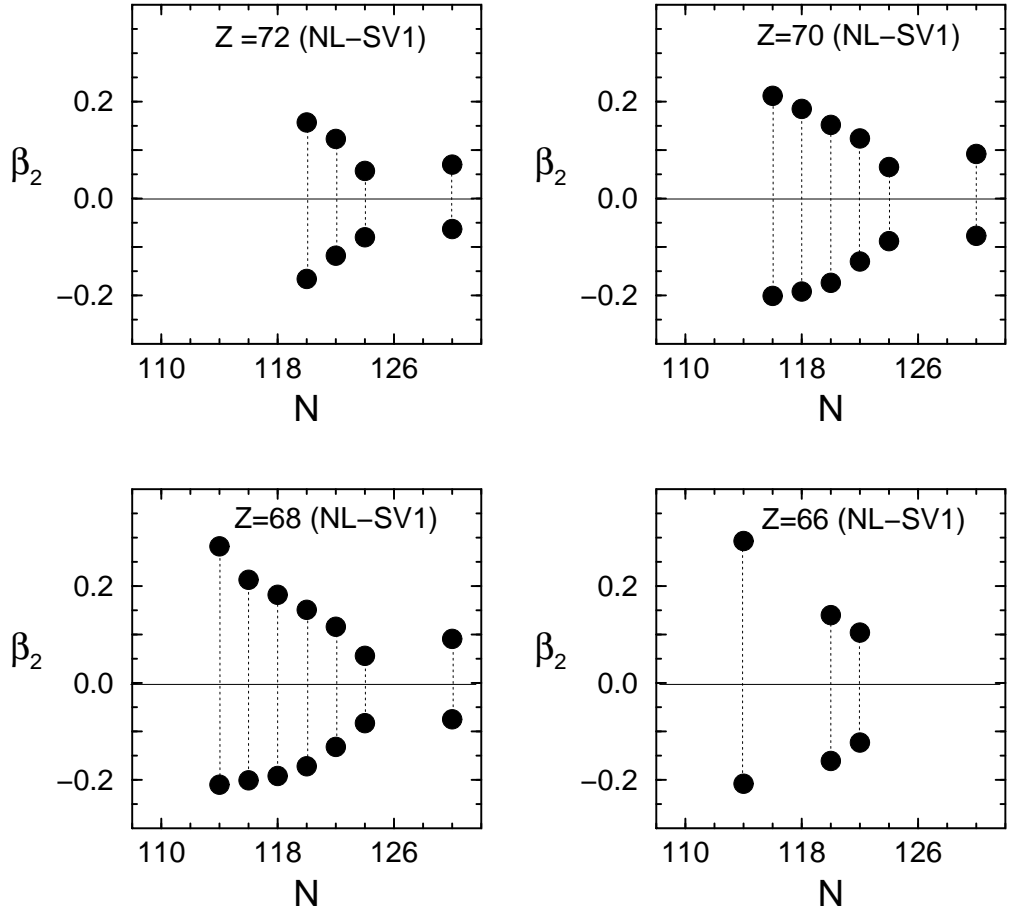}}
\vspace{0cm}       
\caption{The shape-coexistence of prolate and oblate shapes in the 
ground state of nuclei in the isotopic chains with $Z=66-72$.}
\label{fig:4}       
\end{figure}

The shape-coexistence for nuclides in the isotopic chains of $Z=66-72$ are
shown in Fig.~4. Most of the occurrence of the shape-coexistence in these
chains takes place in nuclides below the magic number $N=126$. It is striking
that a significantly large number of nuclides exhibit the shape-coexistence
for the chains $Z=70$ (Yb) and $Z=68$ (Er) and especially those in the latter.
For the Yb chain, the isotopes with $N=116-124$ exhibit the shape-coexistence,
whereas for the Er chain it takes place for the isotopes with $N=114-124$.

The magnitude of deformation for prolate and the corresponding
coexisting oblate shape is comparable and modest. A reduction in 
the value is shown as one approaches the shell-closure.
It is interesting to note that nuclides with $N=124$, which are so close
to the shell-closure, are amenable to a shape-coexistence albeit with
a significantly reduced deformation values. For the isotopic chains 
$Z=72$ (Hf) and $Z=66$ (Dy), in comparison, there are only a limited cases 
of shape-coexistence below $N=126$. In contrast, there is only one case 
each for the shape-coexistence in the 
chains $Z=72$, $Z=70$ and $Z=68$ for nuclides with neutron  
number above $N=126$. For $Z=66$, there is no case of 
shape-coexistence above N=126.

\begin{figure}[h*]
\vspace {0.5 cm}
\hspace{0.5cm}
\resizebox{0.80\textwidth}{!}{%
   \includegraphics{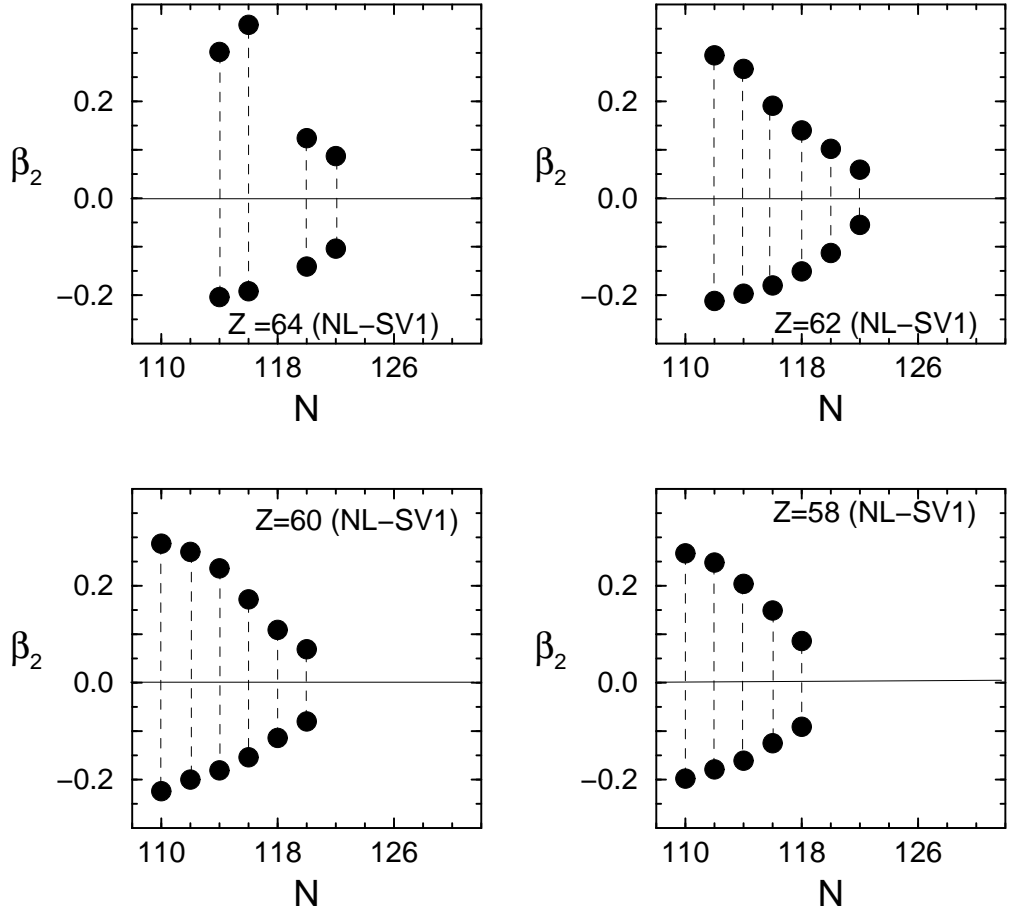}}
\vspace{0cm}       
\caption{The shape-coexistence of prolate and oblate shapes in the 
ground state of nuclei in the isotopic chains with $Z=58-64$.}
\label{fig:5}       
\end{figure}

The shape-coexistence in the isotopic chains $Z=64$ (Gd), $Z=62$ (Sm),
$Z=60$ (Nd) and $Z=58$ (Ce) is shown in Fig.~5. For these chains, there is
no shape-coexistence found for neutron numbers above $N=126$. Just as in the
case of $Z=66$ (Dy) in Fig.~4, its neighbouring chain $Z=64$ (Gd) exhibits
a fewer cases of shape-coexistence as compared to the chains $Z=62$ (Sm) 
and $Z=60$ (Nd) and to some extent $Z=58$ (Ce). The picture for the
shape-coexistence in Fig.~5 is very similar to that in Fig.~4.
Nearly all even-even isotopes in the chains $Z=62$ exhibit
shape-coexistence from $N=112-122$. This is shifted to $N=110-120$ for
the $Z=60$ chain and to $N=110-118$ for the $Z=58$ chain.

There are no cases of shape-coexistence seen for $N > 126$ for any
of the chains in Fig.~5. As noted above, nuclides with $N > 126$ 
are gradually moving towards the neutron drip-line as $Z$ is decreasing
from $Z=64$ to $Z=58$. This is evident also from Fig. 2 on deformations
whereby nuclides with $N \ge 126$ are projected to be spherical.

In summarizing the shape-coexistence in Fig. 4 and 5, it is fair to
say that the phenomenon of shape-coexistence is abound in a large
number of nuclides in the region $N=110-124$ below the magic number.
This implies that the potential energy landscape for nuclides
in this region of the periodic table is relatively 'softer' in
the deformation space. 

\subsection{The two-neutron separation energies}

The shell effects near $N=126$ being the main objective of this work,
we show 2-neutron separation energies $S_{2n}$ in Fig.~6. The $S_{2n}$
values show a plateau like curves for all isotopic chains below 
the magic number $N=126$, with slight undulations here and there due
to modulations caused by the deformation. Yet, most of the curves
show a slight declining trend in going from $N=112$ to $N=126$.
Also, the level of each curve is decreasing as the atomic number
$Z$ of an isotopic chain decreases. 

\begin{figure}[h*]
\vspace {0.5 cm}
\hspace{0.5cm}
\resizebox{0.80\textwidth}{!}{%
   \includegraphics{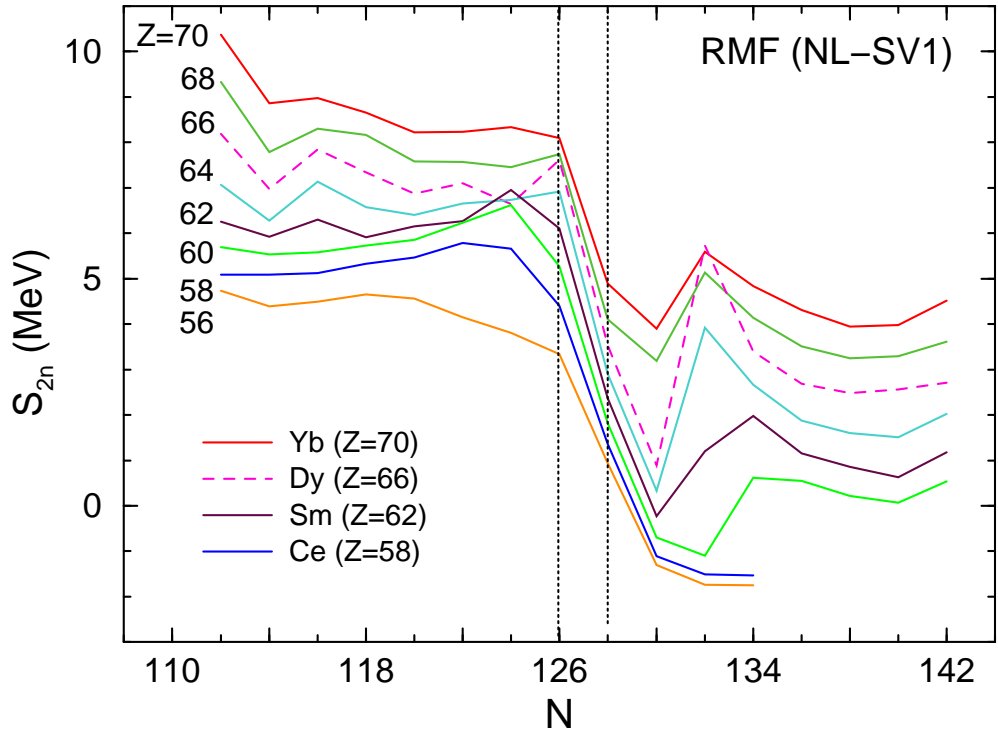}}
\vspace{0cm}       
\caption{The 2-neutron separation energies of nuclei in the 
isotopic chains $Z=56-70$ obtained with the deformed RMF calculations
using the Lagrangian model NL-SV1.}
\label{fig:6}       
\end{figure}

The $S_{2n}$ values demonstrate a sudden decline at $N=128$ for
all the isotopic chains. This is a manifestation of the shell closure
at $N=126$. The decrease in the slope of the curve between $N=126$ and $N=128$
amongst various isotopic chains is minimal, 
implying that the shell gap at $N=126$ does not show a
significant change in going from $Z=70$ to $Z=56$. This means that the shell
gap at $N=126$ remains rather intact even as one approaches the nuclei
near the drip line. There is thus no washing out (quenching) of the 
shell effects at $N=126$ near the neutron drip line. A similar result
was obtained within the framework of the spherical relativistic
Hartree-Bogoliubov approach  in our previous
work (Farhan and Sharma, 2006).

For nuclides with $N=130$ for the isotopic chains $Z=66$ to $Z=56$, there
is a steep decline in the $S_{2n}$ values approaching towards a vanishing
value. This indicates that for $Z<66$ there is an imminent arrival of the
neutron drip line as one goes above $N=126$. 

\subsection{Stability beyond the neutron drip-line}

In going above $N=130$, we observe a spectacular increase in the
$S_{2n}$ values. This is especially the case for the isotopic chains 
Dy ($Z=66$), Gd ($Z=64$) with a peak at $N=132$ and a bit lesser for the
chains of Sm ($Z=62$) and Nd ($Z=60$) showing a peak at $N=134$. 
To some extent, 
this effect is also visible for Yb ($Z=70$) and Er ($Z=68$) with a peak
at $N=132$. The sudden spurt in the $S_{2n}$ value especially for Dy
and Sm isotopes with $N=132$ signals a significantly higher binding 
of additional neutrons as compared to $N=130$ isotopes. This provides
an additional binding energy to an isotope as compared to its lighter
neighbour. Thus, this phenomenon can suitably be termed 
as '{\it stability beyond the neutron drip line}'. In fact,
due to the advent of this feature, the neutron drip line for some
of these isotope chains is extended farther in the neutron space.
To our understanding, this is the first observation of the stability
beyond the neutron drip line. In the subsection below, we will
see as to how this phenomenon is produced.

\begin{figure}[h*]
\vspace {0.5 cm}
\resizebox{0.60\textwidth}{!}{%
   \includegraphics{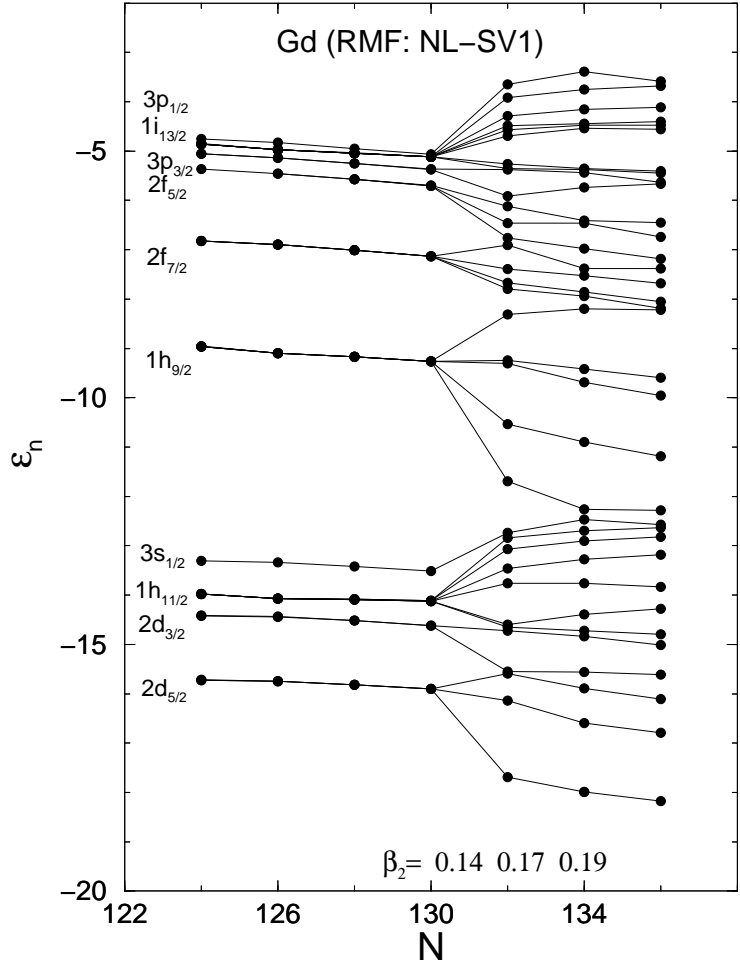}}
\vspace{0cm}       
\caption{The Nilsson single-particle levels for the isotopes
of Gd ($Z=64$) obtained with the Lagrangian model NL-SV1 in
the RMF theory. The isotopes above $N=128$ are deformed as
indicated by the $\beta_2$ values shown at the bottom of the
panel.}
\label{fig:8}       
\end{figure} 

\subsection{The neutron single-particle levels}

In order to visualize as to how the stability beyond the neutron drip line
is produced, we show the neutron single-particle levels (Nilsson levels)
for the isotopes of the Gd ($Z=64$) chain in Fig.~7. The isotopes with
N=124-130 are all spherical and hence there is no splitting of the
$j$-levels. However, in going above $N=130$, the nuclei assume a modest
deformation with $\beta_2 \sim 0.14-0.19$. The effect of this deformation
is clearly evident in the splitting of $j$-levels into various
$\Omega$-orbitals. 

The salient feature of the deformed single-particle
levels in Fig.~7 is that some of the $\Omega$-levels are pushed down
in the energy, thus effectively lowering the 'centre-of-gravity' 
of the neutron $j$-levels for the isotopes with $N=132-136$ as compared
to the isotope with $N=130$. This downward shift in the 
energy levels is then responsible for
the additional binding energy provided by the deformation and thus
bestowing an additional stability beyond the neutron drip line.

\section{SUMMARY AND CONCLUSIONS}
 
We have investigated the nuclear shell effects at the magic number $N=126$ in 
the vicinity of the $r$-process path for the third peak in the $r$-process
abundances. This region passes through the rare-earth region encompassing
$Z=56-72$. Within the framework of the relativistic mean-field theory,
we have calculated the ground-state properties of even-even nuclei in these
isotopic chains. The force NL-SV1 with the nonlinear vector self-coupling
of $\omega$-meson has been employed in this work. 

The results show that a large number of nuclides attain deformation except
those close to the magic number $N=126$. The shape transitions from
oblate-prolate-spherical from $N=110$ to $N=126$ are exhibited by nearly all 
the chains with the point of transitions shifting slightly. We observe
that a large number of nuclei especially those below $N=126$ exhibit
the phenomenon of shape-coexistence of prolate and oblate shapes in
the ground state. 

Analyzing the 2-neutron separation energy of nuclei, it is seen that
the shell gap at $N=126$ does not show much softening in going from
the region away from the drip line towards the neutron drip line. 
Thus, the shell effects at $N=126$ remains rather strong even in the
vicinity of the drip line. This is consistent with an earlier work
(Farhan and Sharma, 2006) wherein nuclear shell effects were studied
within the framework of the relativistic Hartree-Bogoliubov theory
with spherical shapes. As nuclei near and at the magic number are all
spherical as seen in the present work, the result on the shell effects
in the present work in the deformed RMF theory is not expected 
to differ from the earlier work (Farhan and Sharma, 2006).

The most interesting aspect that has emerged from this work is the
increase in the 2-neutron separation energy of nuclides near $N=132-134$
in several isotopic chains and especially for Gd ($Z=64$) and Dy ($Z=66$)
as compared to their lighter counterparts. It is shown that these
nuclei exhibit the phenomenon akin to a stability beyond the neutron drip
line. The analysis of the deformed single-particle levels for these
nuclides shows that due to the splitting of levels caused by the deformation,
several $\Omega$-levels are pushed down in energy, thus lowering down
the single-particle levels effectively. This lends an additional binding 
energy to nuclei though these nuclei are very close to the neutron drip 
line.

\newpage

\hspace{5cm} {REFERENCES}

{\bf Burbidge, E.M., Burbidge, G.R.,  Fowler, A.A.$\&$ Hoyle, F. (1957).} 
Synthesis of elements in stars. Rev. Mod. Phys. \textbf{29}: 547-650.

\vglue0.1cm

{\bf Boguta, J and Bodmer, A.R.  (1977).}  Relativistic calculations of
nuclear matter and the nuclear surface. Nucl. Phys. \textbf{A292}: 413-428.

{\bf Dillmann I. {\it et al.} (2002).} N=82 shell quenching of the classical
r-process waiting-point nucleus $^{130}$Cd. Phys Rev. Lett. \textbf{91}: 
162503.

{\bf  Dobaczewski, J., Flocard, H. $\&$ Treiner, J.  (1984).} 
Hartree-Fock Bogoliubov description of nuclei near the neutron-drip line.  
Nucl. Phys. A {\bf 422}:103-139.   

\vglue0.1cm

{\bf Farhan, A.R. $\&$ Sharma, M.M. (2006).} Strength of nuclear shell 
effects at $N=126$ in the r-process region.  Phys. Rev. {\bf C73}: 045803 
(13 pages).

\vglue0.1cm

{\bf Gambhir, Y.K., Ring, P, $\&$ Thimet, A. (1990).} Relativistic mean-field
theory for finite nuclei. Ann. Phys (N.Y.) {\bf 198}: 132-179. 

\vglue0.1cm

{\bf Gorska, M. {\it et al.} (2009).} Evolution of the N=82 shell gap 
below $^{132}$Sn inferred from core excited states in $^{131}$In.  
Phys. Lett. B {\bf 672}: 313-316. 

\vglue0.1cm

{\bf Hillebrandt, W. (1978).} The rapid neutron-capture process and 
the synthesis of heavy and neutron rich elements. Space Sci. Rev., 
\textbf{21}: 639-702.

\vglue0.1cm

{\bf  Jungclaaus, A. {\it et al.} (2007).} Observation of isomeric
 decays in the $r$-process waiting-point nucleus $^{130}$Cd$_{82}$. 
Phys. Rev. Lett. {\bf 99}:132501 (5 pages).

\vglue0.1cm

{\bf Kratz, K.L., Bitouzet, J.P.,  Thielemann, F.K., M\"oller, 
P. $\&$  Pfeiffer, B. (1993).}  Isotopic $r$-process abundances and nuclear
structure far from stability : Implications for the r-process mechanism.
Astrophys. J. \textbf{403}: 216-238.

\vglue0.1cm

{\bf  M\"oller, P. $\&$ Nix, J. (1992).}  Nuclear
pairing models. Nucl. Phys. \textbf{A536}: 20-60.

\vglue0.1cm

{\bf  M\"oller, P.,  Nix, J.,  Swiatecki, W. (1994).} Nuclear ground-state
masses and deformations.  At. Data Nucl. Data Tables \textbf{59}: 185-381.

\vglue0.1cm 

{\bf  Pearson, J.M., Nayak, R.C. $\&$ S. Goriely. (1996).} Nuclear 
mass formula with Bogoliubov-enhanced shell quenching: application 
to r-process.  Phys. Lett. B \textbf{387}: 455-459.

\vglue0.1cm

{\bf  Pfeiffer, B.,  Kratz, K.L. $\&$ Thielemann, F.K. (1997).} 
Analysis of the solar system r-process abundance pattern with the 
new ETFSI-Q mass formula. Z. Phys.  \textbf{A357}: 235-238.

\vglue0.1cm

{\bf Pfeiffer, B., Kratz, K.L.,  Thielemann, F, -K. $\&$  Walters, 
W.B. (2001).} Nuclear structure studies for the astrophysical $r$-process.  
Nucl. Phys. \textbf{A693}: 282-324.

\vglue0.1cm

{\bf  Rodriguez, T.R.,  Egido, J.L. $\&$  Jungclaus, A. (2008).} 
On the origin of the anomalous behaviour of 2$^+$ excitation energies 
in the neutron-rich Cd isotopes. 
Phys. Lett. B {\bf 668}: 410-413.

\vglue0.1cm

{\bf Serot, B.D.  $\&$  Walecka, J.D.  (1986).} The relativistic 
nuclear many-body problem. Adv. Nucl. Phys. \textbf{16}: 1.

\vglue0.1cm

{\bf Sharma, M.M., Farhan, A.R. $\&$  Mythili, S.  (2000).} 
Shell effects in nuclei with vector self-coupling of the 
$\omega$-meson in the relativistic Hartree-Bogoliubov theory.  
Phys. Rev. C \textbf{61}: 054306 (15 pages)

\vglue0.1cm

{\bf  Sharma, M.M. $\&$ Farhan, A.R.  (2002).} Nuclear shell effects 
near the $r$-process path in the relativistic Hartree-Bogoliubov theory.  
Phys. Rev. C \textbf{65}: 044301 (8 pages)

\vglue0.1cm

{\bf Strutinsky, V.M. (1968).} Shells in deformed nuclei.  Nucl. Phys. 
\textbf{A122}:1-33.

\vglue0.1cm

\end{document}